\documentclass[12pt]{article}
\usepackage{graphicx}
\usepackage{pdfpages}
\usepackage{hyperref}
\usepackage{amssymb}
\usepackage{amsmath}
\usepackage{color}
\usepackage{caption}
\usepackage[normalem]{ulem}
\usepackage[numbers,sort&compress]{natbib}

\newcommand{\be}{\begin{equation}}
\newcommand{\ee}{\end{equation}}

\newcommand{\ben}{\begin{eqnarray}\displaystyle}
\newcommand{\een}{\end{eqnarray}}

\begin{document}

\begin{center}
{\Large 
\bf{When Strings Surprise}
}

\vspace{6mm}

\textit{Nissan Itzhaki$^{1,2,3}$ and Uri Peleg$^{2}$}
\break \break
 $^1$  Department of Physics, Princeton University, Princeton, NJ 08544, USA\\\vspace{1mm}
 $^2$ School of Physics and Astronomy, Tel Aviv University, Ramat Aviv, 69978, Israel\\\vspace{1mm}
 $^3$School of Natural Sciences, Institute for Advanced Study\\\vspace{1mm}
 1 Einstein Drive, Princeton, NJ 08540, USA\\
\end{center}

\vspace{5mm}

\begin{abstract}

We argue that on-shell excitations with large {\it negative} energies are created rapidly when the string coupling increases with time.
This does not indicate an inconsistency in string theory since the negative energy on-shell excitation is always entangled with an on-shell excitation with a positive energy. The total energy of this energy-EPR state vanishes. We discuss the reason the energy-EPR states appear in string theory and the role they might play in black hole physics. 
\end{abstract}


\vspace{5mm}

%

\newpage

It is fair to say that thirty years after Joe Polchinski posed the question \cite{Polchinski:1994mb} there is still confusion about what string theory is.
This short note aims to add to the confusion.

We claim that in soft backgrounds ($\alpha^{\prime}R\ll1$ and $g_s =e^{\Phi} \ll 1$)  in which the string coupling grows slowly with time, on-shell excitations with large negative energies are rapidly created.  These excitations do not imply inconsistency since they do not appear by themselves. An on-shell excitation with large negative energy is always created together with another on-shell excitation with positive energy such that the total energy vanishes. Schematically, the  state takes the form
\be\label{one}
|\Psi\rangle=\int dE_1 |E_1\rangle \otimes |E_2=-E_1\rangle.
\ee
Namely, we argue that 
string theory admits energy-EPR states where the total energy (rather than total spin in \cite{BEPR,Bell}) vanishes. 
The vacuum in quantum field theory is filled with pairs of this form. The difference is that here the excitations are on-shell.  This means, in particular, that an observer can take her time in making a precise measurement of the energy of, say, the first excitation and obtain a negative value, $E_1<0$.

We start by considering the simplest  background in which $\partial_{\mu} \Phi$ is time-like and points to the future: a time-like linear dilaton background
\be 
ds^2 = - dt^2 + dx^2 +dy^2 +dz^2+ \mbox{Compact}, \quad \Phi = Q t. \label{eq:TLD}\ee
We work with $\alpha^{\prime}=1$ and take  $Q>0$, so that the string coupling constant, $g_s=e^{Q t}$, blows up in the future. We consider $Q\ll 1$ which means that, at least naively,  a low-energy effective description is expected to be valid. As we shall see the creation of the energy-EPR states occurs long before approaching the strong coupling region at large $t$.
To illustrate the existence of states like (\ref{one}) we need only one non-compact space-like direction. We take the number of non-compact space-like directions to be three to emphasize that this could happen in our world.\footnote{As we shall see the process that leads to (\ref{one}) is a local process that requires only
that locally $\partial_t \Phi >0$. Hence measuring an on-shell excitation with a large negative energy does not mean we are heading towards a singularity in the future. }

\begin{figure}
\includegraphics[width=11cm]{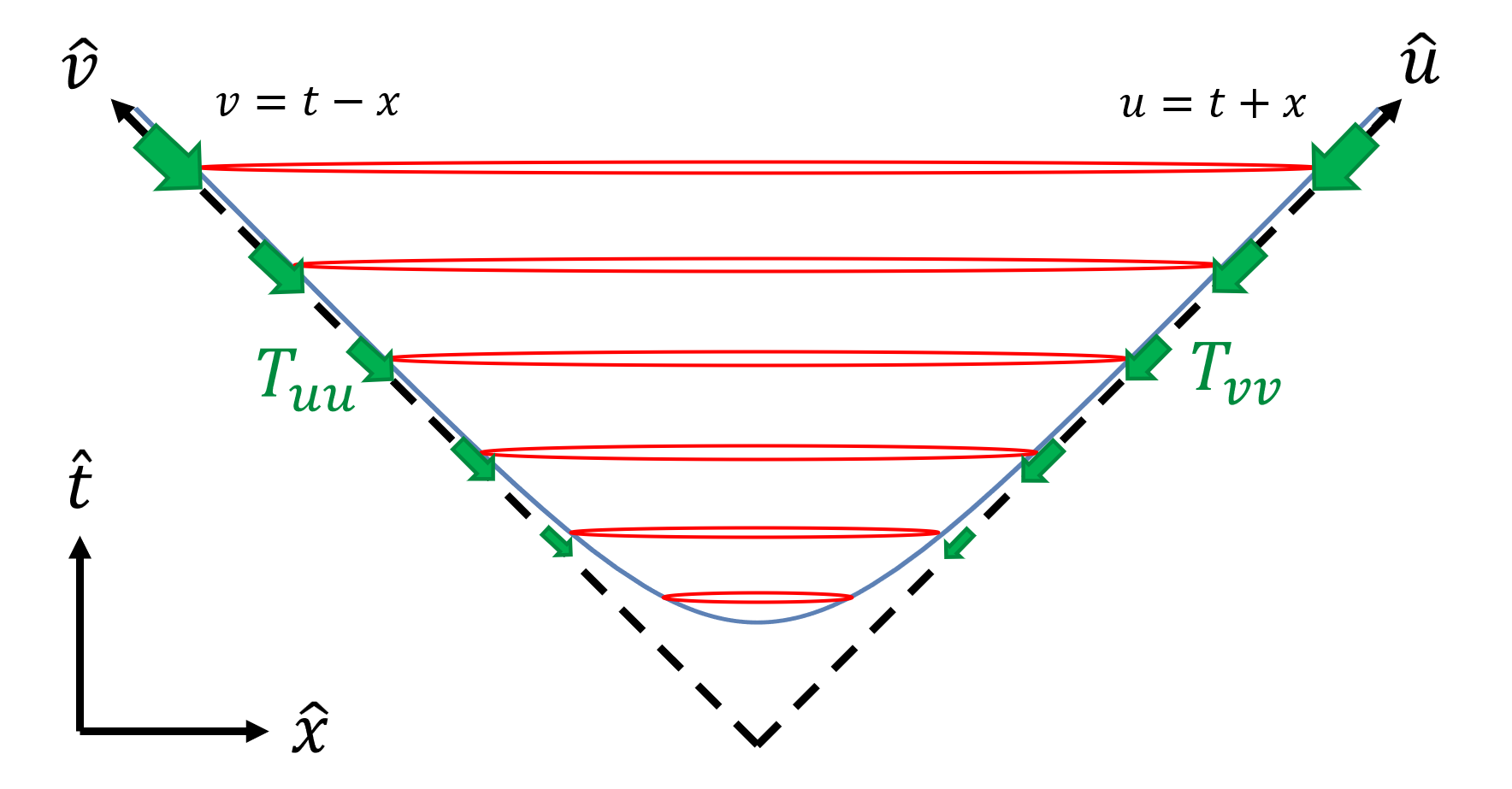}
\centering
\captionsetup{width=\linewidth}
\caption{The IFS solution. 
The green arrows represent the null energy flux at the fold, pointing backward in time to indicate negative energy. 
The flux becomes more negative over time. This feeds energy into the bulk of the IFS, allowing it to grow and become macroscopic.
}
\label{fig:IFS}
\end{figure}

For any positive  $Q$,  no matter how small, this background admits classical string solutions to the equation of motions and Virasoro constraints that are absent for $Q=0$ \cite{instability inside the BH}
\be t(\sigma,\tau) = t_0 +  Q\ln \left(\frac{1}{2}\cosh \left(\frac{\sigma}{ Q}\right) + \frac{1}{2}\cosh\left(\frac{\tau}{ Q}\right)\right), \label{ifss}\ee
and $x=x_0+\sigma,~y=y_0,~z= z_0$
with $-\infty < \sigma, \tau, < \infty.$ 
This solution describes a closed folded string that is created at $t=t_0, ~ x=x_0,~ y=y_0$, and $z=z_0$. The string expands rapidly and asymptotically the fold travels at the speed of light. See figure \ref{fig:IFS}.

At the fold, $\tau=0$, we have $\partial_{\tau}t=0$. Therefore, the same solution in the upper half-plane ($-\infty<\sigma<\infty, ~0\leq \tau<\infty$) with a Neumann boundary condition at $\tau=0$ describes an Instant Open String (IOS)\footnote{We thank I. Klebanov for this observation.}. Like the fold of the IFS, the endpoints of the IOS travel faster than light.

The origin of the discontinuity in $Q$ -- the fact that a solution exists for any $Q>0 $ and does not exist for $Q=0$ -- is the following.  By definition,  $\alpha'$ corrections associated with the target space curvature, $H$ field, and second derivatives cannot dominate, in soft backgrounds, the leading Virasoro constraints, 
\be\label{ol}
g_{\mu\nu}\partial_{\pm}X^{\mu}\partial_{\pm}X^{\nu}.
\ee  
 The dilaton gradient, however, is different. It is formally subleading in the $\alpha'$ expansion, but since it contributes to the Virasoro constraints a linear term in $X^{\mu}$,
\be\label{ya}
\alpha'\partial_{\mu}\Phi \partial_{\pm}^2 X^{\mu},
\ee
it can dominate (\ref{ol}) for a short (world sheet) time, even if $\partial_{\mu}\Phi$ is small. This is what allows the string to fold. Near the fold (\ref{ya}) dominates (\ref{ol}), and away from the fold (\ref{ol}) dominates. There is a sense in which a time-like dilaton gradient violates the equivalence principle. It triggers the creation of light strings that are simply absent in its absence.\footnote{Space-like  dilaton gradient also leads to solutions for any $Q\neq 0$ that are absent when $Q=0$ \cite{Maldacena:2005hi}. The difference is that these solutions have an infinite energy so they do not affect the low-energy dynamics.}

One might argue that the existence of instant strings  (IFSs and/or IOSs) is not that surprising. The background in which they are created is time-dependent, and it is often the case that time dependence leads to the creation of states from the vacuum. Hence it is natural to wonder if this is yet another stringy version of the Schwinger mechanism \cite{Schwinger:1951nm} in which the role of the electric field is played by $Q$.\footnote{The open string Schwinger mechanism was discussed in \cite{Bachas:1992bh}, and a closed string analog,   in which the $H$ field plays the role of the electric field, was discussed in \cite{Dowker:1995sg}. } There are, in fact, some crucial differences between instant string creation and the Schwinger mechanism. We find it useful to discuss these differences as they emphasize the unique features of instant strings and why they decay into (\ref{one}). 

The first difference is the creation scale. In the Schwinger mechanism, the creation scale grows as the electric field is decreased. In particular, the creation scale blows up as the electric field vanishes.  Consequently calculating basic quantities, such as the production rate, is challenging for a varying electric field. For instant strings, the situation is the opposite. As is clear from (\ref{ifss}) the creation scale is $Q$. Hence as we decrease $Q$ the instant string creation process becomes more and more local. Thus as long as the curvature and the second derivatives of $\Phi$ are small they cannot affect local properties of the  IFS. 

This locality argument  implies that the instant strings are not related to the perturbative  tachyons that usually appear in 
(\ref{eq:TLD}) since the length scale associated with the creation of these tachyons, $1/Q$, is much larger than the scale associated with the creation of instant strings, $Q$.
Indeed a small second derivative of $\Phi$ can render the tachyon massless (or massive), but, as discussed above,  it will have little effect on the instant strings.

The locality of the IFS creation has important implications also for the IFS production rate.
In the time-like linear dilaton background (\ref{eq:TLD}) the IFS production rate is \cite{Worldsheet Description of IFS} 
\be\label{ks}
\Gamma_{IFS}\sim \frac{Q^2}{g_s^2},\ee
where $g_s$ is the string coupling at $t_0$.
The locality argument above implies that in a more general background, with a time-like $\partial_{\mu}\Phi$ that points to the future, it is
\be\label{od}
\Gamma_{IFS}\sim \frac{(\partial_{\mu}\Phi)^2}{ g_s^2}. 
\ee
Moreover, the instant string solution is expected to be well approximated by (\ref{ifss}) at distances shorter than the curvature scale. 

The fact that $\Gamma_{IFS}\sim 1/g_s^2$ follows from the fact that IFSs are created at the classical level, and it deserves some discussion. 
In particular, (\ref{od}) implies that in the time-like linear dilaton background (\ref{eq:TLD}) 
the production rate blows up in the far past, so the assumption that the space is empty is wrong. 

This, in fact, can be tested in a rather precise way.
A production rate that scales like $1/g_s^2$ should leave its mark on the sphere partition function.
In particular, if the initial state at  $t=t_i$ is the vacuum, $| 0(t_i)\rangle$, then the amplitude to remain in the vacuum at a later time $t_f$ is exponentially suppressed due to IFS production
\be\label{hv}
\langle 0 (t_i)| 0(t_f)\rangle =\exp\left( -V \int_{t_i}^{t_f}  \Gamma_{IFS}(t) ~ dt\right)\sim \exp\left(-V Q e^{-2 Q t_i}\right),
\ee
where $V$ is the volume and we used (\ref{ks}). We assume here that $t_f-t_i\gg Q$, which implies that the integral is dominated by $t_i$. We also take $Q\ll 1$ which means that the dilute IFS-gas approximation is valid and that interactions among the IFSs can be neglected. In this case (\ref{hv}) should be compared with the time-like linear dilaton sphere partition function.

In string theory, however,  $t_i$ and $t_f$ cannot be finite. We must take $t_i=-\infty$ and $t_f=\infty$, which gives $0$. A way to put an effective cutoff at $t_i$ in string theory is to add a Liouville potential. In space-like linear dilaton both options for the Liouville wall ($\exp(b\phi)$ and $\exp(\phi/b)$) cut off the strong coupling region. In the time-like case, we can either cut the strong coupling region (in the future, in our case) or the weak coupling region. For this particular calculation, we have to cut off the weak coupling region since there $\Gamma_{IFS}$ blows up, and assume that the singularity at the future does not matter for this calculation since there $\Gamma_{IFS}$ vanishes.

As usual (see e.g. \cite{Polchinski's Book}) the full partition function  $Z_{vac}$ is related to the single string partition function, $Z_1$, via 
\be
Z_{vac}=\exp(Z_1).
\ee
Thus, for  $Q\ll 1$,  the time-like Liouville $Z_1$ should be compared with 
\be\label{goku}
Z_1=- Q e^{-2Q t_i}.
\ee
It appears that an agreement with IFS consideration requires that $Z_1$ is real and negative. This  is not standard for the partition function in a time-like direction, which usually is imaginary.  

Time-like Liouville theory, however, is not a standard theory. In particular, its relation to space-like Liouville, via an analytic continuation of $b$, is rather subtle \cite{Zamolodchikov:2005fy,Harlow:2011ny}.
Luckily, using the Coulomb gas approach, $Z_1$ was calculated in time-like Liouville by Giribet \cite{Giribet:2011zx},
who found  
\be\label{gi}
Z_1= \frac{(1+b^2)(\pi\Lambda \gamma(-b^2))^{Q/b}}{\pi^3 Q \gamma(-b^2)\gamma(-b^{-2})}.
\ee
KPZ scaling \cite{Knizhnik:1988ak} relates the $(\pi\Lambda \gamma(-b^2))^{Q/b}$ with the $e^{-2Qt_i}$ in (\ref{goku}). The comparison we are left with is between the $-Q$ in (\ref{goku}) and the factor of $(1+b^2)/Q\gamma(-b^2)\gamma(-b^{-2})$ in (\ref{gi}), which,  for $Q\ll 1$, indeed agree,
up to a numerical factor that we cannot determine at the moment. To check the numerical factor one needs to calculate $\Gamma_{IFS}$ in the presence of the Liouville wall.

The second difference is that in the Schwinger effect, the role of the electric field is twofold. It triggers the pair creation, and it also feeds the pair with energy after the creation. It accelerates the electron, say, to the left while accelerating the positron to the right. 
In the instant string case, $Q$ is only the trigger for its creation, but it does not feed it with energy after the creation. The instant string feeds itself. Namely, the instant string is created from the vacuum with zero energy (and momentum), and,
 as apparent from the existence of the zero modes $t_0$ and $x_0$, the total energy and momentum remain zero at later times.  This was verified in a direct calculation  \cite{KN}. As expected, away from the fold the only non-vanishing component of the energy-momentum tensor is $T_{uv}$, (with $u=t+x$ and $v=t-x$) which in the small $Q$ limit takes a particular simple form\footnote{The exact energy-momentum tensor can be found in \cite{KN}. }
\be
T_{uv} = \frac{1}{2\pi\alpha'}\Theta(u)\Theta(v),
\ee
associated with the tension of the folded string. Energy momentum conservation fixes $T_{uu}$ and $T_{vv}$ at the fold to be
\be  T_{uu} = -\frac{1}{2\pi\alpha'}\Theta(v)v \delta(u), \quad T_{vv} = -\frac{1}{2\pi\alpha'}\Theta(u)u  \delta(v), \label{LimitEMT}\ee
which implies that at the fold there is a negative null flux. 
Evidently, the way the instant string feeds itself is by transferring energy from the folds toward the bulk of the string allowing it to grow with time. 

\begin{figure}
\includegraphics[width=11cm]{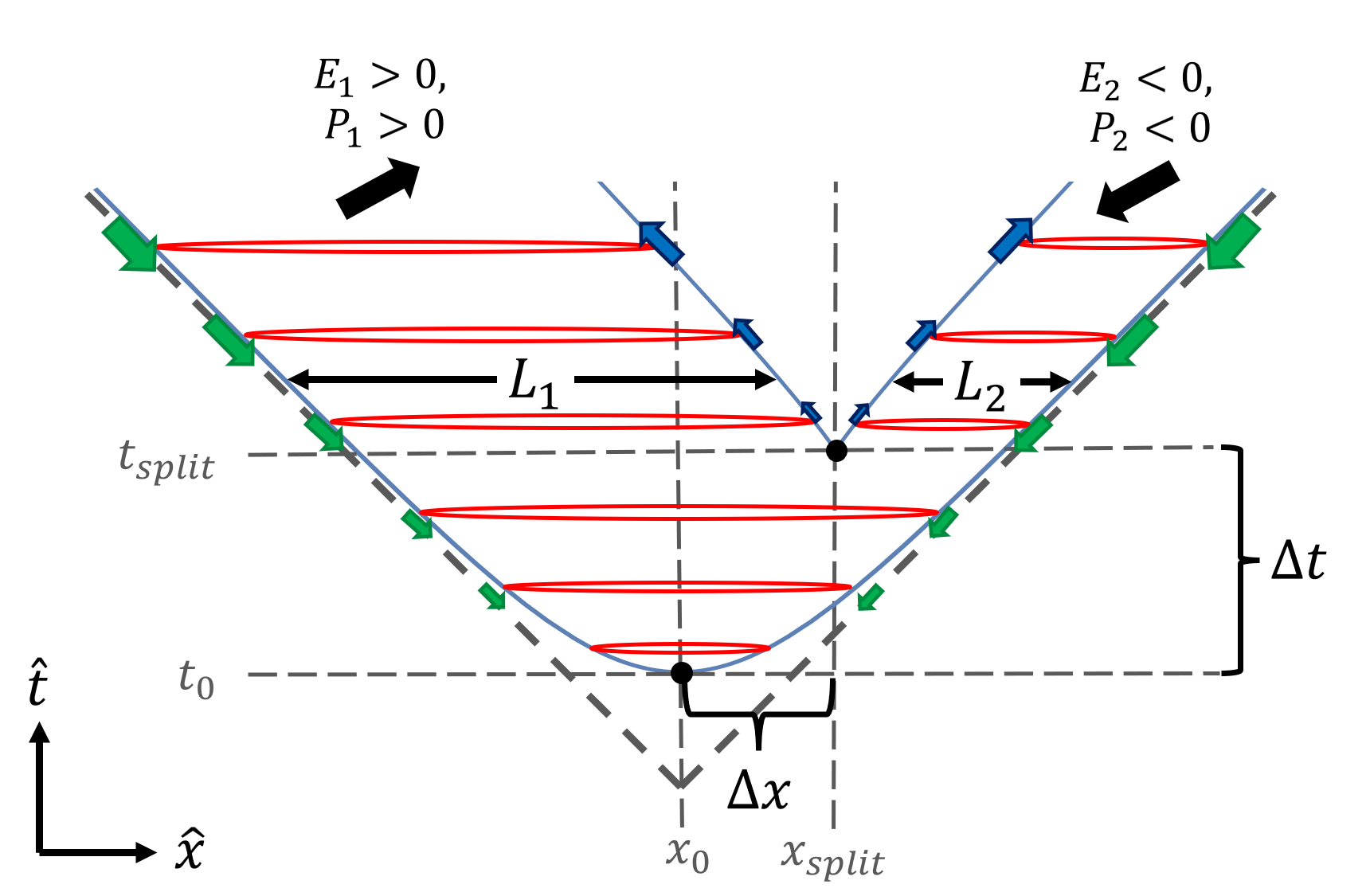}
\centering
\captionsetup{width=\linewidth}
\caption{IFS decay, the offset of the breaking point from the center, $\Delta x$, determines the distribution of the bulk energy between the two components, leading to $E_1=-E_2=\frac{\Delta x}{2\pi \alpha'}$. The total momentum of each component is due to the folds $P_2=-P_1=\frac{\Delta t}{2\pi\alpha'}$. The widths of each component's bulk are respectively $L_1=\Delta t + \Delta x$ and $L_2=\Delta t - \Delta x$. The blue and green arrows at the folds in each time slice represent the energy and momentum associated with the inner and outer folds.}
\label{fig:IFS split}
\end{figure}

The instant string solution is quite unusual. On the one hand, it describes a light state with $E=P=0$. On the other hand, even from afar, it does not look at all like a particle. In particular, it becomes macroscopic at late times, and so, at finite string coupling, it can split. 
The IFS splits into two folded strings (see Figure \ref{fig:IFS split}), and the IOS splits into two open strings.

Since the total momentum and energy of an instant string vanish we have 
\be\label{78}
E_2=-E_1,~~~~P_2=-P_1.
\ee
If the splitting takes place right in the middle of the instant string then $E_1=E_2=0$.  If the splitting occurs at some point to the right (left)  of the middle point then $E_2=-E_1 > (<) 0$.
The splitting of an instant string is a local process that does not depend on the location of the splitting (as long as it is away from the fold). Hence the wave function of the two strings associated with an instant string that splits at time $\Delta t$ after its creation is well approximated by   
\be\label{21}
|\Psi(\Delta t)\rangle\sim \int_{-\frac{\Delta t}{2\pi\alpha'}}^{\frac{\Delta t}{2\pi\alpha'}} d E_1 \left|E_1=\Delta x/2\pi \alpha^{\prime}, P_1=-\Delta t/2\pi\alpha' \right\rangle \otimes\left|E_2=-E_1, P_2=-P_1 \right\rangle, 
\ee
which can be viewed as an energy-EPR state. Note that one of the strings always has a negative energy and that this negative energy is typically quite large. It is of the order of $\Delta t/\alpha'$. Using \cite{Dai:1989cp} we can estimated that $\Delta t\sim l_s/g_s$  and so $E_1\sim M_s/g_s$. 

Although the state $\left|E_1, P_1 \right\rangle$ has fixed quantum numbers, and, unlike the IFS, from afar it does look like a particle, it cannot be described by a $(1,1)$ vertex operator (even if $E_1>0$). The reason is that while the total energy and momentum associated with this state do not vary in time there is quite a bit of dynamic involved in the time evolution of $\left|E_1, P_1 \right\rangle$. The simplest way to see this is to consider the energy-momentum tensor associated with 
$\left|E_1, P_1 \right\rangle$. This state has two folds (or ends, in the case of the IOS). One inherited from the instant string and another due to the splitting. Causality implies that the fold that was inherited from the instant string is not aware of the splitting. Hence the negative null flux at the fold is still, say, 
\be  T_{uu} = -\frac{1}{2\pi\alpha'}\Theta(v)v \delta(u), \ee
and, in particular, it decreases with time (it becomes more negative). 
The bulk of the string still contributes $T_{uv} = \frac{1}{2\pi\alpha'}$, and so by energy-momentum conservation at the new fold there is a positive null flux 
\be T_{uu} = \frac{1}{2\pi\alpha'}(v-L_2)\Theta(v-L_2) \delta(u-L_1), \ee
with $L_1 =  \Delta t + \Delta x$ and  $L_2 =  \Delta t - \Delta x,$ that grows with time (see figure (\ref{fig:IFS split})).
The mechanism by which  $\left|E_1, P_1 \right\rangle$ evolves in time is by transferring energy from the fold inherited from the IFS to the new fold through the bulk of the string.

For  $L_1\gg l_s$, we expect $\left|E_1, P_1 \right\rangle$ to split further. It is natural to expect the splitting to stop when $L_1\sim l_s$. This suggests that the final state associated with the decay of an instant string involves two states, inherited from the fold of the IFS, with negative energy of the order of $-M_s/g_s$. They point in opposite directions and are entangled with many soft modes with positive energy. The total energy of this energy-EPR state vanishes.

We would like to end with some questions:\\
$\bullet$ What are the possible imprints of the energy-EPR states? In Cosmological scenarios that involve a time-dependent dilaton, these states are created but they do not contribute to the average time evolution. The main contribution in Cosmology is due to the IFSs (before they decay) that induce negative pressure at no energy density cost \cite{Itzhaki:2021scf}.  The implication of this will be discussed elsewhere \cite{ta}. The energy-EPR states do appear to be relevant for fluctuations. It should be interesting to study the differences between the fluctuations associated with the energy-EPR states and standard cosmological fluctuations and see if there is a sharp prediction that can be made. 

Another possibility is a direct detection of an on-shell excitation with negative energy. 
Unfortunately,  at least in the IFS decay case, the negative energy excitation appears to couple only via gravity to the standard model fields which makes detection unrealistic. 
Note that since the energy is negative, the gravitational shock wave produced by such an excitation induces time advance which could lead to causality violation. 
To violate causality we need, however, to have several such excitations and control their production location and momenta. This does not appear to be an easy task given the way they are produced.

$\bullet$ Why do these energy-EPR states appear in string theory? 
A possible answer is related to the fact that the dilaton determines the amount of classical or coarse-grained entropy via
$
G_N^{-1} \sim e^{-2\Phi}. 
$
As a result when the dilaton varies in time so does the classical entropy. This appears to be the source of the radiation of quantum or fine-grained entropy in the form of energy-EPR states. The IFS appears to play the role of a convertor as it converts the coarse-grained entropy into a fine-grained entropy.

If we define
$ \Psi=e^{-\Phi},$
we have that the coarse-grained entropy scales like  $\Psi^2$ and (\ref{od}) implies that $(\partial \Psi)^2$ determines the fine-grained entropy production. It is, therefore, natural to dub $\Psi$ an entropon. The term "entropon" appears in the condensed matter literature in the context of active solids \cite{entropons}, which are solids that involve self-propelled excitations. Amusingly, there seems to be some analogy between active matter and string theory with time-dependent dilaton. Standard closed string modes are the analog of phonons. Both are excitations that are present even when the solid is inactive.
When the dilaton grows with time string theory becomes active. It includes new self-propelled excitations: the instant strings that grow by feeding themselves. Their decay products, the energy-EPR states, dominate the entropy production. 

$\bullet$ Are there implications to  Black Holes? The BH in which it is easiest to address this is the near extremal NS5-branes \cite{Maldacena:1997cg}, i.e. the 2D BH \cite{Witten:1991yr,Mandal:1991tz,Elitzur:1990ubs,Dijkgraaf:1991ba}. The region behind the horizon of such a BH includes a time-like $\partial_{\mu}\Phi$ that points to the future. In \cite{Worldsheet Description of IFS} it was shown that the production rate (\ref{od}) implies that the number of IFSs an infalling observer encounters on the way to the singularity is of the order of the BH entropy. Here we claim that an IFS decays into an energy-EPR state, which means that the Bekenstein-Hawking entropy associated with near extremal NS5-branes is of the order of the fine-grained entropy associated with the energy-EPR states that are created inside the BH. Combining this with \cite{Maldacena:2013xja}, assuming that the energy-EPR state also forms a wormhole, we seem to conclude that near extremal NS5-branes are filled with tiny wormholes. In the context of JT gravity, a related claim was made in \cite{Stanford:2022fdt}. Since the energy-EPR states involve excitations with negative energy, the nature of these wormholes, if they exist, is likely to be nonstandard. 

\vspace{6mm}
{\it Acknowledgements:} We thank 
D. Gross, A. Hashimoto, G. Horowitz, V. Hubeny, J. Minahan, I. Klebanov, H. Ooguri, M. Rangamani, and  A. Sen for discussions. Work supported in part by the ISF (grant number 256/22). This research was supported in part by grant NSF PHY-2309135 to the Kavli Institute for Theoretical Physics (KITP).

\end{document}